\documentstyle[11pt,paspconf,epsf]{article}

\begin{document}

\title{A southern VLA-based gravitational lens search}
\author{Joshua N. Winn, Jacqueline N. Hewitt, Paul L. Schechter}
\affil{Massachusetts Institute of Technology, 77 Massachusetts Avenue, Cambridge, MA 02139}

\begin{abstract}
We present a status report on our search
for multiple-image QSOs among southern
radio sources. Our goal is to identify new lenses for
use in studies of cosmology and galaxy structure.
To date we have examined
3300 radio sources, identified one probable lens and
one probable binary QSO along with about 50 candidates
deserving further follow-up.
\end{abstract}


\section{Introduction}

We are searching for galaxy-scale gravitational
lenses among southern radio sources, with the
hope that new lenses will advance theories
of galaxy structure (by modeling
lens potentials) and cosmology (by measuring
lensing rates or differential time-delays).
We chose the region $\delta = 0^{\circ}$ to
$-40^{\circ}$, because this region is relatively
unexplored for lenses (see figure on next page)
and well-situated for new
southern observatories, and yet is (just) visible
to the VLA and VLBA. We exclude the region within
$10^{\circ}$ of the galactic plane to simplify
optical follow-up.

\begin{figure}
\plotone{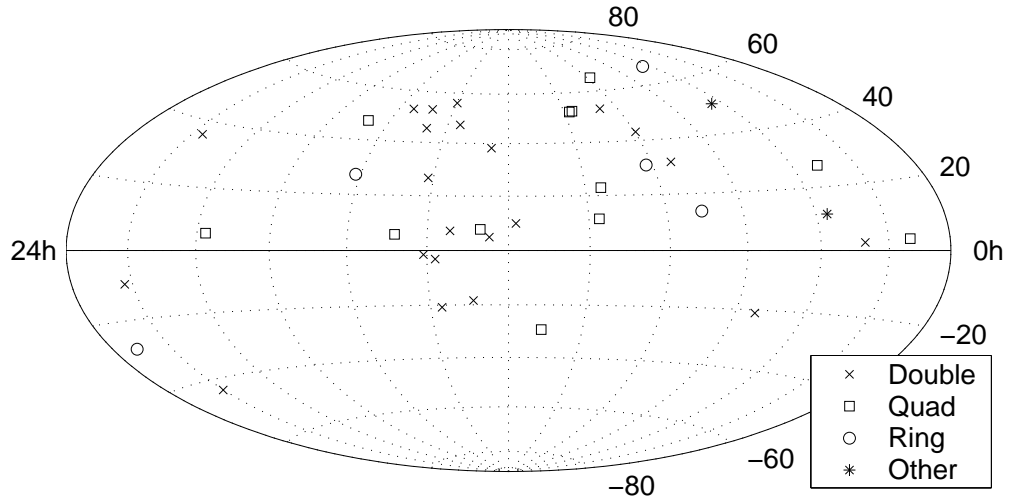}
\caption{Aitoff plot of well-established cases
of multiple-image QSOs, showing the relative scarcity of
southern lenses. Data courtesy of CASTLES survey
(http://cfa-www.harvard.edu/castles).}
\end{figure}

Our strategy is similar to the
one adopted previously by CLASS,
a successful northern lens search program
(see Browne; Myers; Rusin; this volume).
To screen large numbers of radio sources, we used
the VLA in its A-array to
obtain 30-second snapshots at 8.4 GHz of
thousands of objects selected from the Parkes-MIT-NRAO (PMN)
catalog of southern radio sources.

We selected sources that are flat-spectrum ($\alpha > -0.5$, where
$S_{\nu} \sim \nu^{\alpha}$), since these sources tend to be
core-dominated.
This makes mapping easier to automate, and makes cases
of lensing easier to recognize.
The auxiliary catalogs we used to compute spectral indices were
the the Parkes (2.8 GHz), NVSS (1.4 GHz), Molonglo (408 MHz), and
Texas (365 MHz) catalogs.

\section{Progress}

To date we have acquired about 3300 snapshots. There are about 50 objects
exhibiting multiple compact components separated by $0.2$ to $6.0$
arcseconds, which will undergo (as appropriate):
multiwavelength VLA imaging (to check if the components have similar
spectral indices), VLBA imaging, and direct optical imaging.	

Objects that pass these tests are candidates for
spectroscopy (to check if the components have similar features and
redshifts) and HST imaging (to disentangle the QSO images and
the lens galaxy).
None of our objects have made it to this final stage yet, although in
two cases we have identified objects that are double point-sources
in both radio and optical images, with the same separation and
orientation. In one of these cases, VLA follow-up revealed that the
spectral indices of the components are very similar,
indicating a probable lens. In the other case the components
have different spectral indices, suggesting a binary QSO. Follow-up
with the VLBA is being scheduled.

\section{Extending the sample}

By combining our observations with the prior observations of
A. Patnaik and (separately) A.B. Fletcher \& B.F. Burke,
we have assembled a nearly-complete
sample of about 5000 flat-spectrum radio sources satisfying the criteria:
\begin{eqnarray} 
S_{4.8 \mathrm{GHz}} \geq 60 {\mathrm m}{\mathrm J}{\mathrm y} &
{\mathrm f}{\mathrm o}{\mathrm r} & 0^{\circ} \geq \delta \geq -30^{\circ}
\nonumber \\
S_{4.8 \mathrm{GHz}} \geq 80 {\mathrm m}{\mathrm J}{\mathrm y} &
{\mathrm f}{\mathrm o}{\mathrm r} & -30^{\circ} > \delta \geq -40^{\circ} 
\nonumber
\end{eqnarray}

Once it has been thoroughly mined for lenses, this sample can be
used to measure the lensing rate, thereby
constraining the cosmological constant.
Based on predicted and empirically measured lensing rates (for example
by the similar, northern lens survey CLASS) we expect ultimately to
identify 5-10 new lenses.

\acknowledgments

J.N.W. thanks the Fannie and John Hertz foundation for financial support.
We also acknowledge the support of NSF grants AST96-17028 and AST96-16866.

\end{document}